\documentclass[journal]{IEEEtran}

\pdfoutput=1

\usepackage{amsmath,bm,amssymb,amsfonts,amsthm}
\usepackage[linesnumbered, ruled, vlined]{algorithm2e}
\usepackage{graphicx}
\usepackage{xcolor} 
\usepackage{relsize}
\usepackage{textcomp}
\usepackage{listings}
\usepackage{gensymb}

\usepackage{changepage}
\usepackage{hhline}
\usepackage{multirow}
\usepackage{nomencl}
\usepackage{mathtools}
\usepackage{commath}
\usepackage{booktabs}
\usepackage{subfiles}
\usepackage{tabularx}
\usepackage{lscape}
\usepackage{rotating}

\usepackage{verbatim}
\usepackage{comment}

\usepackage[normalem]{ulem}
\usepackage[utf8]{inputenc}
\usepackage[hyphens]{url}
\usepackage{longtable}
\usepackage{lineno}
    \modulolinenumbers[5]
\usepackage[caption=false]{subfig}
\usepackage{wrapfig}
\usepackage{enumitem}
    \setlist{nolistsep}
\usepackage[hidelinks]{hyperref}
\usepackage{makecell}
\usepackage{multicol}
\usepackage{colortbl}

\usepackage{scalerel}
\usepackage{tikz}
\usetikzlibrary{svg.path}
\definecolor{orcidlogocol}{HTML}{A6CE39}
\tikzset{
  orcidlogo/.pic={
    \fill[orcidlogocol] svg{M256,128c0,70.7-57.3,128-128,128C57.3,256,0,198.7,0,128C0,57.3,57.3,0,128,0C198.7,0,256,57.3,256,128z};
    \fill[white] svg{M86.3,186.2H70.9V79.1h15.4v48.4V186.2z}
                 svg{M108.9,79.1h41.6c39.6,0,57,28.3,57,53.6c0,27.5-21.5,53.6-56.8,53.6h-41.8V79.1z M124.3,172.4h24.5c34.9,0,42.9-26.5,42.9-39.7c0-21.5-13.7-39.7-43.7-39.7h-23.7V172.4z}
                 svg{M88.7,56.8c0,5.5-4.5,10.1-10.1,10.1c-5.6,0-10.1-4.6-10.1-10.1c0-5.6,4.5-10.1,10.1-10.1C84.2,46.7,88.7,51.3,88.7,56.8z};
  }
}
\newcommand\orcidicon[1]{\href{https://orcid.org/#1}{\mbox{\scalerel*{
\begin{tikzpicture}[yscale=-1,transform shape]
\pic{orcidlogo};
\end{tikzpicture}
}{|}}}}

\usepackage[noadjust]{cite}

\begin{document}

\title{\huge A Review of the GIC Blocker Placement Problem}

\author{
    Arthur~K.~Barnes $^{1}$\orcidicon{0000-0001-9718-3197},
    Adam~Mate $^{1}$\orcidicon{0000-0002-5628-6509},
    and Russell~Bent $^{1}$\orcidicon{0000-0002-7300-151X}
    \vspace{-0.20in}

\thanks{Manuscript submitted:~Oct.~09,~2023.
Current version: Mar.~20,~2024.
}

\thanks{$^{1}$ The authors are with the Advanced Network Science Initiative at Los Alamos National Laboratory, Los Alamos, NM 87545 USA. \\ Email: abarnes@lanl.gov, amate@lanl.gov, rbent@lanl.gov.}

\thanks{LA-UR-23-24544. Approved for public release; distribution is unlimited.}

}

\markboth{IEEE/IAS 60th Industrial \& Commercial Power Systems Technical Conference, May~2024}{}

\maketitle


\begin{abstract}
Space weather poses a tremendous threat to power systems: geomagnetic disturbances could result in widespread disruptions and long-duration blackouts, including severe damage to system components.
To mitigate their impacts, a handful of strategies exist, with the most promising being the deployment of transformer neutral blocking devices. The high cost of these devices, however, precludes their installation at all substations; this motivates the development of effective solutions for the cost-effective placement of such devices.
While the current state-of-the-art in blocker placement methods is insufficient to be applied to real-sized power grids, ongoing research continues to increase the size of networks for which the placement problem remains tractable.
Along these lines, the contributions of this paper are two fold: first, a comprehensive overview of the current state-of-the-art in blocker placement methods is provided; and second, a complete optimization formulation -- implemented and benchmarked in an open-source software -- for the blocker placement problem is presented.
\end{abstract}


\section{Introduction} \label{sec:introduction}
\indent
\vspace{-0.1in}

The bulk electric system (BES) faces risks on many different time-scales -- both short- and long-term events, varying based on the amount of warning time present before the threat and the time expected to restore service after the threat -- which can lead to major system disruptions \cite{NA-resilience-2017}.
From a grid resilience perspective, the most consequential and hardest to predict risks are called high-impact low-frequency events (HILFs) \cite{NERC-HILF-2010}.
HILFs are a challenge for protecting the BES because they may cause widespread, long duration blackouts -- potentially involving transient instability issues, cascading outages -- or severe damage to grid components, such as large power transformers, which are challenging and time-consuming to repair \cite{mate21-pmsgmd-cascade, mate21-pmsgmd}.

The North American Electric Reliability Corporation identifies geomagnetic disturbances (GMDs) as HILFs \cite{NERC-HILF-2010}: these solar-driven incidents disrupt the Earth's magnetic field and drive geomagnetically induced currents (GICs) in the conductive infrastructure, and may be predicted a few hours or few days before the threat.
Since they are rare, there is a lack of empirical data on how the BES will respond to such threats; GMDs change the physics of grid operation, so power-flow formulations must be modified \cite{juvekar2021gic}.
Therefore, accurately simulating their impact and developing cost-effective mitigation strategies that can operate on their time-scales is a complex research challenge.
\vspace{0.1in}

\vspace{-0.1in}
\subsection{Approaches to Mitigate GMD Impacts}
\indent \vspace{-0.15in}

The research community has developed approaches over the years to help plan for and mitigate the potential impacts of GMDs on the BES.
Considered mitigation strategies include load-shedding \cite{overbye2022towards}, transmission line switching \cite{kazerooni2017mitigation, lu2017optimal, ryu2022mitigating}, generation switching (re-dispatch) \cite{overbye2022towards}, shunt/series-capacitor switching \cite{zhang2022machine}, and addition of transformer neutral (dc-current) blocking devices \cite{yang2019optimal}.

Line switching reduces the total transmission capacity of the power grid, and likely requires generation re-dispatch, customer load-shedding, or both to achieve a new generation-demand balance.
Generation switching, removing generation from the grid, breaks GIC flow via disconnecting generator step-up transformers.
Transformer neutral blocker placement -- the installation of GIC dc-current blocking devices -- consists of injecting a shunt capacitor in series with the grounding point of transformer neutrals; this does not interfere with the power-carrying capacity of the grid, but may interfere with ground-fault relaying, which in turn requires additional switching devices leading to increased capital costs.

This paper focuses on the placement of transformer neutral blocking devices to mitigate GMD impacts.
Given the high cost of these devices, sparse placement is necessary; it is an unrealistic expectation to install such devices at all transformer neutrals in all substations. 
In addition, the literature has suggested that the number of transformers that experience high GICs during GMD events is also sparse \cite{horton_magnetohydrodynamic_2017}. 
Finally, the placement itself is a challenging problem because installing a blocker at one transformer can increase GICs at other transformers throughout the power grid; therefore, simple approaches, such as greedy methods, are likely unsuitable in this case.
These three factors motivate the need for optimization-based placement methods to guide the choices of where to place blocking devices.

\renewcommand{\arraystretch}{1.15}%
\begin{table*}[!htbp]
\caption{Summary of GIC Blocker Placement Methods}
\begin{center}
\vspace{-0.05in}
\begin{tabular}{cccrrcr}
\hline
Problem Formulation & Objective & Solver & Busses & Runtime & CPU & Authors \\ 
\Xhline{2\arrayrulewidth}
dc GIC flow & min. $I_{eff}$ & GA & ~100 & -- & -- & \cite{ning2019research} \\
dc GIC flow & min. $I_{eff}$ & GA & ~110 & -- & -- & \cite{yang2019optimal} \\
time-extended dc GIC flow & avg./peak $q_{loss}$ min. over time & SA & -- & -- & -- & \cite{liang2019optimal} \\
SDP, relaxed placement & min. total $q_{loss}$ or max. $q_{loss}$ & -- & 20 & -- & -- & \cite{zhu2014blocking} \\
SOCP, relaxed placement, lin. $q_{loss}$ & -- & -- & -- & -- & -- & \cite{zhu2016blocking} \\
lin. sensitivity factors, line switching & -- & greedy & -- & 0.29 s & Core i7 & \cite{kazerooni2017mitigation} \\
scen.-based lin. $q_{loss}$, ac power flow & min. blocker count & GA & ~1,000 & -- & -- &  \cite{shengquan2022optimal} \\
-- & min. blocker count & surrogate opt. & 118 & -- & -- & \cite{lamichhane2022optimal} \\
MINLP, ac power flow & -- & GA & -- & 6 h & Core i7-2600 & \cite{etemadi2014optimal} \\
MINLP, ac power flow & -- & branch-and-cut & 1,875 & 2:47 h & Core i5 & \cite{liang2015optimal} \\
MINLP, no ac power flow & -- & SA & 1,875 & 48 h & 3.10GHz Core i5  & \cite{liang2019optimal} \\
MINLP, harm. ac power flow, xfmr heating & -- & -- & 118 & 25 h & 12 cores & \cite{rezaei2018optimal} \\
\hline
\end{tabular}
\end{center}
\label{table:lit-review}
\vspace{-0.25in}
\end{table*}
\renewcommand{\arraystretch}{1}%

\vspace{-0.15in}
\subsection{Blocker Placement for GMD Mitigation}
\indent
\vspace{-0.15in}

GICs are quasi-dc currents -- produced by the low-frequency electric fields that GMDs induce on the Earth's surface -- that appear in the conductive infrastructure and flow into the high-voltage network through the neutrals of transformers \cite{pirjola2000-gic}.
They are dependent on system characteristics (geographical location of substations, transformer parameters), geomagnetic source fields (amplitude, frequency content, spatial characteristics), and the Earth conductivity structure (substation grounding resistance, influence on geoelectric fields) \cite{shetye2015-gmd, boteler2017modeling}.

Placement of blocking devices on transformer neutrals can break the flow of GICs and block them from entering/exiting the electrical grid at grounded points.
However, dc-current blocking devices do not prevent induced currents on transmission lines themselves, and only drive the ``effective'' GICs ($I_{eff}$) to zero for auto-transformers and transformers with a single grounded winding (typically delta-grounded-wye generator step-up transformers); these devices have limited benefits for transformers with multiple grounded windings, such as grounded-wye-grounded-wye or three-winding transformers.
Additionally, even with complete dc-blocker placement, it is not guaranteed that $I_{eff}$ will be zero across the entire system, nor that the placement problem will be feasible \cite{boteler2017modeling}; alternate mitigation methods, such as series capacitor placement or line switching, should also be explored.

The literature on blocking device placement can generally be categorized using two criteria: the problem formulation, and the method used to solve the problem.
Given the computational complexity associated with modeling the physics of power flows, existing research has considered a wide range of problem formulations, such as complete non-linear formulations \cite{etemadi2014optimal, liang2015optimal, rezaei2018optimal, liang2019optimal, lamichhane2022optimal}, mathematical relaxations \cite{zhu2014blocking, zhu2016blocking, kazerooni2017mitigation} and linearized approximations \cite{ning2019research, yang2019optimal, liang2019optimal, shengquan2022optimal}.
Some work has focused only on formulating the equivalent dc network for calculating GIC flows \cite{ning2019research, yang2019optimal}, others on the dc network with reactive power loss ($q_{loss}$) calculations \cite{zhu2014blocking, liang2019optimal} for coupling with ac power flows, and finally, on formulations that feature the fully coupled dc and ac networks \cite{etemadi2014optimal, zhu2016blocking, rezaei2018optimal, kazerooni2017mitigation, shengquan2022optimal}.
Other potential additions include transformer hot-spot constraints \cite{rezaei2018optimal}, multi-time-series extension \cite{liang2019optimal, mate21-pmsgmd}, and various objectives; often the objective is to minimize total or maximum $I_{eff}$ \cite{ning2019research, yang2019optimal}, minimize total or maximum $q_{loss}$ \cite{liang2019optimal}, minimize maximum transformer hot-spot temperature, minimize load-shedding, or minimize dc-blocker placement cost.
Solution methods may rely on linear solvers, local solvers, or genetic algorithms/simulated annealing \cite{ning2019research, yang2019optimal}.

While research have tried to tackle the scalability of this problem, current state-of-the-art is limited to test cases with fewer than $2\times10^3$ busses, far less than the size of real-world BES networks, such as the U.S. Eastern Interconnect with $6\times10^4$ busses.
The rest of this paper presents a complete optimization formulation for the GIC dc-blocker placement problem -- which includes features of practical network models such as line charging or switched shunts -- and highlights challenges in obtaining a globally optimum solution in reasonable time.
A reference implementation for this formulation is provided in PowerModelsGMD.jl\footnote{https://github.com/lanl-ansi/PowerModelsGMD.jl} \cite{mate21-pmsgmd}, which is benchmarked on a set of publicly available test cases.

\section{Problem Formulation} \label{prob_form}
\indent
\vspace{-0.1in}

This section describes the developed optimization-based problem formulation for the strategic placement of blocking devices, based on full ac power flow equations.
Relative to conventional ac transmission network problems -- such as load flow, optimal power flow, optimal line switching, minimum load-shedding -- a positive-sequence representation of the problem is insufficient; the blocker placement problem requires the addition of a separate quasi-dc representation of the transmission network, which is coupled to the ac network through a set of constraints associated with transformers.
The quasi-dc representation itself is fairly similar to the zero-sequence representation of the ac network, with two major distinctions: inductances are neglected and assumed to be short circuits, while capacitances are neglected and assumed to be open circuits.
It is an architectural choice for implementation whether the dc network is to be modeled as a completely separate network, or added to the ac network as a set of special dc busses (nodes) and lines (branches or edges).

In contrast to traditional ac problems, additional data is required for the blocker placement problem:
ground electric field magnitude and direction, which are assumed to be known;
geographical locations of transmission line endpoints, or the full path of lines for increased fidelity;
resistance to remote earth of substation grounding grids; and
transformer core configurations.
Furthermore, it is necessary to explicitly model generator step-up transformers, which are frequently not included in published transmission network datasets.

\vspace{0.05in}
The end-to-end workflow of the GIC blocker placement problem is as follows:
\begin{enumerate}
\item generation of the dc network given the ac network;
\item line coupling calculations to determine the induced voltages along transmission lines;
\item calculations of the ``effective'' GICs ($I_{eff}$) on transformers;
\item calculations of the transformer reactive power losses ($q_{loss}$); and
\item solving the ac power flow problem.
\end{enumerate}

\vspace{0.05in}
Depending on the exact problem to be solved, some of these steps may be coupled within the same optimization problem rather than being solved sequentially.

\newpage
\subsection{Objective Function}
\indent
\vspace{-0.15in}

A number of objective functions for guiding the choice of GIC blocker placement are reasonable, including:
\begin{itemize}
\item minimizing the sum-squared of the $I_{eff}$ (subject to constraints on the number of blockers placed);
\item minimizing the total cost of placing GIC dc-blockers (subject to power flow constraints and a limit on the amount of allowed load-shedding); or
\item minimizing the amount of load-shedding (subject to constraints on the number of blockers placed).
\end{itemize}

\vspace{0.05in}
There is a clear trade-off between power system resilience (as expressed by the total amount of load served during a GIC event) and the total number of blockers placed throughout the system.
Additional objective functions may also be considered:
the total of the weighted sum of $I_{eff}$ ($I_{eff}$ is a weighted sum of the transformer winding currents, which is proportional to the core magnetic flux density);
the total of the weighted sum of the cost of placing GIC dc-current blockers; or
the total cost of operating generation (however, as HILF events are considered, fuel cost of generation is arguably not a concern).

\vspace{0.05in}
The cost associated with operating generation, typical for optimal power flow problems under nominal conditions, is as follows:

\begin{equation}
\allowdisplaybreaks
\small
\mathrm{min.} \smashoperator{\sum_{i \in \mathcal{G}}} \kappa_i^0 + \kappa_i^1 p_i^g + \kappa_i^2 (p_i^g)^2
\label{eq:opf-obj}
\end{equation}
\noindent where
$\mathcal{G}$ is the set of generators in the ac network;
$p_i^g$ is the power output of generator $i$; and 
$\kappa_i^0,\,\kappa_i^1,\,\kappa_i^2$ are the cost coefficients of generator $i$.

\vspace{0.1in}
\subsubsection{Minimize cost of sum-squared GIC}
\indent

The cost associated with minimizing GIC throughout the system is:
\begin{equation}
\allowdisplaybreaks
\small
\mathrm{min.} \smashoperator{\sum_{i \in \mathcal{E}^d}} |I_i|^2
\label{eq:gic-obj}
\end{equation}
\noindent where
$\mathcal{E}^d$ is the set of directed edges in the dc network, oriented from dc node $i$ to dc node $j$; and
$I_i$ is the $I_{eff}$ on edge $i$ of that network.

\vspace{0.1in}
\subsubsection{Minimize cost of blocker placement}
\indent

The cost associated with placing GIC dc-current blockers is:
\begin{equation}\allowdisplaybreaks
\small
\mathrm{min.} \smashoperator{\sum_{i \in \mathcal{N}^{db}}} \kappa_i^b z_i^b
\label{eq:dc-blocker-placement-obj}
\end{equation}
\noindent where
$\mathcal{N}^{db}$ is the set of nodes in the dc network that correspond to transformer neutrals and are candidates for placed blockers;
$\kappa_i^b$ is the cost of placing a blocker at node $i$; and
$z_i^b \in \{0,1\}$ is the binary placement variable for the blocker in node $i$.

\vspace{0.1in}
\subsubsection{Minimize amount of load shed}
\indent

The cost associated with load-shedding is:
\begin{equation}
\allowdisplaybreaks
\small
\mathrm{min.} \smashoperator{\sum_{i \in \mathcal{D}}} \abs{p_i^d} \kappa_i^d z_i^d
\label{eq:load-shed-obj}
\end{equation}
\noindent where
$\mathcal{D}$ is the set of loads in the ac network;
$p_i^d$ is the dispatched power of load $i$; 
$\kappa_i^d$ is the cost of shedding load $i$; and
$z_i^d \in \{0,1\}$ is the load-shedding variable corresponding to load $i$.

\newpage

\vspace{-0.15in}
\subsection{AC Power Flow Constraints}
\indent
\vspace{-0.15in}

The ac power flow constraints consist primarily of the set of equations that represent the physics of power flow on an ac transmission network, and the set of inequalities that represent the operational limits of the network.

\vspace{0.1in}
\subsubsection{Slack bus constraint}
\indent

To provide symmetry breaking, $\theta_i$ voltage angles are constrained to be zero:

\noindent for $ \forall \; i\in \mathcal{N}^{as} $
\begin{equation}
\allowdisplaybreaks \footnotesize
\theta_i = 0
\label{eq:slack-bus-constr}
\end{equation}
\noindent where
$\mathcal{N}^{as}$ is a set of slack nodes in the ac network.

\vspace{0.1in}
\subsubsection{Nodal power balance constraint}
\indent

\vspace{0.05in}
\noindent for $ \forall \; i\in {\mathcal{N}^a} $
\begin{subequations}
\allowdisplaybreaks \footnotesize
\begin{align}
\smashoperator{\sum_{j \in {\mathcal{E}_i^{a+}}} } p_j^+ - \smashoperator{\sum_{j \in {\mathcal{E}_i^{a-}}} } p_j^-
= \sum_{j \in \mathcal{G}_i} p_j^g z_j^g - \sum_{j \in \mathcal{D}_i} p_j^d z_j^d - v_i^2 \sum_{j \in \mathcal{S}_i} g_j^s z_j^s
\\
\smashoperator{\sum_{j \in {\mathcal{E}_i^{a+}}} } q_j^+ - \smashoperator{\sum_{j \in {\mathcal{E}_i^{a-}}} } q_j^-
= \sum_{j \in \mathcal{G}_i} q_j^g z_j^g - \sum_{j \in \mathcal{D}_i} q_j^d z_j^d - v_i^2 \sum_{j \in \mathcal{S}_i} b_j^s z_j^s - \sum_{j \in \mathcal{L}_i} q_j^{loss}
\label{eq:ac-nodal-balance-cons}
\end{align}
\end{subequations}
\noindent where
$\mathcal{N}^a$ is the set of nodes in the ac network;
$\mathcal{E}_i^+$ is the set of directed edges connected to node $i$ that are oriented towards node $i$ in the ac network, while $\mathcal{E}_i^-$ is the set of directed edges connected to node $i$ that are oriented against node $i$;
$p_j^+$ and $q_j^+$ are the real and reactive power flowing out the end of edge $j$, respectively, while $p_j^-$ and $q_j^-$ are the real and reactive power flowing into the end of edge $j$, respectively;
$\mathcal{G}_i$ is the set of generators in the ac network connected to node $i$, $p_j^g$ is the real power output of generator $j$ and $q_j^g$ is the reactive power output of generator $j$, and $z_j^g \in \{0,1\}$ is the binary on-off variable associated with generator $j$;
$\mathcal{D}_i$ is the set of loads in the ac network connected to node $i$, $p_j^d$ is the real power demand of load $j$ and $q_j^d$ is the reactive power demand of load $j$, and $z_j^d \in \{0,1\}$ is the binary on-off variable associated with load $j$;
$v_i$ is the voltage magnitude at node $i$, while $q_j^{loss}$ is the reactive power losses associated with transformer primary windings connected to node $i$;
$\mathcal{S}_i$ is the set of shunts in the ac network connected to node $i$, $g_j^s$ is the real portion of the admittance of shunt $j$, while $b_j^s$ is the reactive portion of the admittance of shunt $j$, and $z_j^s \in \{0,1\}$ is the binary on-off variable associated with shunt $j$; and
$\mathcal{L}_i$ is the set of $q_{loss}$ pseudo-loads -- associated with the GMD-induced reactive power loss for transformers -- in the ac network connected to node $i$, while $q_j^{loss}$ is the GMD-induced reactive power loss associated with transformer $j$ connected to node $i$.

\vspace{0.1in}
\subsubsection{Branch flow constraint}
\indent

Constraints for edge $i$, connecting nodes $j$ and $k$:

\noindent for $ \forall \; i \in {\mathcal{E}^a} $
\begin{subequations}
\allowdisplaybreaks \footnotesize
\begin{align}
p_i^- &=  -g_i v_j v_k \cos(\theta_j-\theta_k) + b_i v_j v_k \sin(\theta_j - \theta_k) + g_i v_j^2
\label{eq:ac-branch-flow-cons-pij}
\\
q_i^- &= -b_i v_j v_k \cos(\theta_j-\theta_k) + g_i v_j v_k \sin(\theta_j - \theta_k) + \left(b_i + \frac{1}{2}b_i^{sh}\right) v_j^2
\label{eq:ac-branch-flow-cons-qij}
\\
p_i^+ &= g_i v_j v_k \cos(\theta_j-\theta_k) - b_i v_j v_k \sin(\theta_j - \theta_k) - g_i v_k^2
\label{eq:ac-branch-flow-cons-pji}
\\
q_i^+ &= b_i v_j v_k \cos(\theta_j-\theta_k) - g_i v_j v_k \sin(\theta_j - \theta_k) - \left(b_i + \frac{1}{2}b_i^{sh}\right) v_k^2
\label{eq:ac-branch-flow-cons-qji}
\end{align}
\label{eq:ac-branch-flow-cons}
\end{subequations}
\noindent where 
$\mathcal{E}^a$ is the set of directed edges in the ac network;
impedances are represented in rectangular coordinates, where $g_i$ and $b_i$ are the real and reactive portions of the series admittance of edge $i$, respectively, while $b_i^{sh}$ is the shunt admittance of edge $i$ resulting from line capacitance; and
voltages are represented in polar coordinates, where $v_j$ is the magnitude of the voltage at node $j$, and $\theta_j$ is the angle of the voltage at node $j$.

\vspace{-0.15in}
\subsection{AC Operational Limit Constraints}
\indent
\vspace{-0.15in}

Voltage limits at each node are:

\noindent for $ \forall i \in {\mathcal{N}^a} $
\begin{equation}
\allowdisplaybreaks \footnotesize
\underline{v}_i \leq v_i \leq \overline{v}_i
\label{eq:ac-voltage-bounds}
\end{equation}
\noindent where
the lower and upper bounds of the voltage magnitude $v_i$ at node $i \in \mathcal{N}^a$ are set by the range $[\underline{v}_i,\,\overline{v}_i]$.

\vspace{0.1in}
Thermal limits of edges in both directions are:

\noindent for $ \forall i \in {\mathcal{E}^a} $
\begin{subequations}
\allowdisplaybreaks \footnotesize
\begin{align}
(p_i^-)^2 + (q_i^-)^2 \leq \overline{s}_i^2 \\
(p_i^+)^2 + (q_i^+)^2 \leq \overline{s}_i^2 
\end{align}
\label{eq:ac-branch-limit}
\end{subequations}
\noindent where
$\overline{s}_i$ is the thermal limit of edge $i$.

\vspace{0.1in}
Voltage stability limits on the phase angle difference on edge $i$, between nodes $j$ and $k$, are enforced:

\noindent for $ \forall i \in {\mathcal{E}^a} $
\begin{equation}
\allowdisplaybreaks \footnotesize
\underline{\theta}_i \leq \theta_j - \theta_k \leq \overline{\theta}_i 
\label{eq:ac-voltage-angle-limit}
\end{equation}
\noindent where
$\underline{\theta}_i$ and $\overline{\theta}_i$ represent the upper and lower bounds on the phase angle difference.

\vspace{0.1in}
The capacity of power generation:

\noindent for $ \forall i \in \mathcal{G} $
\begin{subequations}
\allowdisplaybreaks \footnotesize
\begin{align}
\underline{p}_i^g &\leq p_i^g \leq \overline{p}_i^g \\
\underline{q}_i^g &\leq q_i^g \leq \overline{q}_i^g
\end{align}
\label{eq:ac-gen-limit}
\end{subequations}
\noindent where 
$\mathcal{G}$ is a set of generators in the ac network connected to node $i \in \mathcal{N}^a$;
$\underline{p}_i^g$ and $\overline{p}_i^g$ represent the lower and upper bounds on real power for generator $i$; and
$\underline{q}_i^g$ and $\overline{q}_i^g$ represent the lower and upper bounds on reactive power for generator $i$.

\vspace{-0.15in}
\subsection{Quasi-DC Voltage and Current Constraints}
\indent
\vspace{-0.15in}

To formulate the nodal current balance constraint, an edge in the dc circuit is linked to an edge in the ac circuit:

\noindent for $ \forall i \in {\mathcal{N}^d} $
\begin{equation}
\allowdisplaybreaks \footnotesize
\smashoperator{ \sum_{j \in \mathcal{E}^{d+}_i}}I_j - \smashoperator{\sum_{j \in \mathcal{E}^{d-}_i}} I_j = (1 - z_i^b)a_i^sV_i
\label{eq:gic-kcl}
\end{equation}
\noindent where
$\mathcal{N}^d$ is the set of nodes in the dc network;
$\mathcal{E}_i^{d+}$ is the set of directed edges connected to node $i$ in the dc network that are oriented towards node $i$, while $\mathcal{E}_i^{d-}$ is the set of directed edges connected to node $i$ that are oriented against node $i$; $I_j$ is the GIC flow through dc branch $j$;
$z_i^b \in \{0,1\}$ is the binary placement variable for the blocker for node $i$, which when placed forces GIC flow into a transformer neutral to 0;
$V_i$ is the quasi-dc voltage at node $i$; and 
$a_i^s$ is the resistance to remote earth at $i$. This constraint introduces a bilinear term with both $z_i^b$ and $V_i$.

\newpage

\vspace{0.1in}
Ohm's law constraint:

\noindent for $ \forall i \in \mathcal{E}^{d} $
\begin{equation}
\allowdisplaybreaks \footnotesize
I_i = z_l a_i(V_j - V_k + \mathbf{E}_i)
\label{eq:gic-ohms-law}
\end{equation}
\noindent where
$l = A_i$ is the ac edge associated with dc edge $i$;
$z_l$ is the status of ac edge $l$; and
$a_i$ is the conductance of dc edge $i$.

\vspace{0.1in}
GIC on a transmission line is determined by the induced voltage source $\mathbf{E}_i$, an unknown real-valued quantity that depends on the displacement of the line in addition to the magnitude and orientation of the ground electric field along that displacement.
While this developed formulation does not address how to handle this uncertainty, existing literature offers solutions; for example, scenario-based approaches or robust optimization approaches could be included \cite{barnes2019resilient, ryu2022mitigating}.

\vspace{0.1in}
The notion of $I_{eff}$ is to calculate a weighted sum of currents through transformer windings that is proportional to the transformer core flux, and therefore to the GIC-induced $q_{loss}$ of the transformer.
This varies for different transformer configurations \cite{overbye2012integration}:

\noindent for $ \forall i \in {\mathcal{E}^a} $
\begin{equation}
\label{eq:gic_effect_3_Idmag}
\allowdisplaybreaks \footnotesize
\widetilde I_i =
\begin{cases}
I_{i^H} & \text{if $i \in \mathcal{E}^\Delta$} \vspace{.1cm} \\
{\frac{\alpha_i I_{i^H} + I_{i^L} }{\alpha_i}} & \text{if $i \in \mathcal{E}^y$} \vspace{.1cm} \\
{\frac{\alpha_i I_{i^S} + I_{i^C} }{\alpha_i + 1}} & \text{if $i \in \mathcal{E}^\infty$} \vspace{.1cm} \\
I_{i^H} + \frac{I_{i^L}}{\alpha_i} + \frac{I_{i^T}}{\beta_i} & \text{if $i \in \mathcal{E}^3$} \vspace{.1cm} \\
0 & \text{otherwise}
\end{cases}
\end{equation}
\noindent where $\mathcal{E}^\Delta$ is the set of delta-delta connected transformers, $\mathcal{E}^y$ is the set of grounded-wye-delta connected transformers, $\mathcal{E}^\infty$ is the set of autotransformers, while $\mathcal{E}^3$ is the set of three-winding transformers;
$\alpha_i = V_{i^H}^b/V_{i^L}^b$ is the transformer turns ratio for the primary and secondary windings, while $\beta_i = V_{i^H}^b/V_{i^3}^b$ is the transformer turns ratio for the primary and tertiary windings for the case of three-winding transformers; $V_{i^H}^b$, $V_{i^L}^b$, and $V_{i^L}^3$ are the nominal voltages for the high-side node, low-side node, and tertiary node corresponding to transformer $i$, respectively;
$\widetilde I_i$ is the $I_{eff}$ current on the dc edge;
$I_{i^H}$ is the high-side winding current and $I_{i^L}$ is the low-side for two- or three-winding transformers;
$I_{i^S}$ is the series winding current and $I_{i^C}$ is the common winding current for auto-transformers; and 
$i^H=H_i^d,\, i^L=L_i^d,\, i^S=S_i^d, \, i^C=C_i^d,\, i^3=T_i^d$ are set index variables for transformer terminal nodes in the dc network.

\vspace{0.1in}
The magnitude of $I_{eff}$:

\noindent for $ \forall i \in \mathcal{E}^{\tau} $
\begin{equation}
\allowdisplaybreaks \footnotesize
\overline{I}_i = \left|{\widetilde{I}_i}\right| 
\label{eq:gic-ieff}
\end{equation}

Last, the $q_{loss}$ increase for a transformer resulting from GIC-induced half-cycle saturation:

\noindent for $ \forall i \in \mathcal{E}^{\tau} $
\begin{equation}
\allowdisplaybreaks \footnotesize
\small
q^{loss}_j = \sqrt{\frac{2}{3}}\cdot\frac{S_i^b}{V_j^b} \cdot K_i\overline{I}_{i}^d v_j
\label{eq:xfmr-reactpwrloss}
\end{equation}
\noindent where
$j = H_i$ is the high-side node associated with transformer $i$;
$q_j^{loss}$ is the GIC-induced reactive power loss at the high-side node $j$;
$S_i^b$ is the base power of transformer $i$;
$V_j^b$ is the rated voltage at the high-side node $j$;
$K_i$ is the GIC reactive power loss coefficient of transformer $i$;
$\overline{I}_{i}^d$ is the absolute value of $I_{eff}$; and
$v_j$ is the high-side node voltage. This constraint also introduces a bilinear term in $\overline{I}_{i}^d$ and $v_j$. While the ac node voltage has relatively tight bounds, typically in the range $[0.9,1.1]$, this is not the case for $I_{eff}$, which is typically in the range $[0,1000]$ A.

\vspace{-0.15in}
\subsection{Blocker Placement Constraints} \label{subsec:blocker-placement-constr}
\indent
\vspace{-0.15in}

To handle the trade-off between the total amount of load served (load lost) during a severe contingency and the total number (cost) of GIC dc-blockers placed throughout the system, additional constraints are introduced.

\vspace{0.1in}
\subsubsection{Cost of dc-current blockers} \label{subsubsec:bpc-cost-blockers}
\indent

To restrict the total cost of blocker placement under a specified threshold budget:
\begin{equation}\allowdisplaybreaks
\small
\smashoperator{\sum_{i \in \mathcal{N}_b^d}} \kappa_i^b z_i^b \le \overline{c}_b
\label{eq:dc-blocker-const-cons}
\end{equation}
\noindent where 
$\kappa_i^b$ is the cost of placing a GIC dc-blocker at node $i$; and
$\overline{c}_b$ is the total budget for blocker placement.

\vspace{0.1in}
This constraint is intended to be used in conjunction with a minimum load-shedding objective.

\vspace{0.1in}
\subsubsection{Number of dc-blockers} \label{subsubsec:bpc-num-blockers}
\indent

To restrict the total number of GIC dc-blockers placed in the system under a specified threshold:

\begin{equation}\allowdisplaybreaks
\small
\smashoperator{\sum_{i \in \mathcal{N}^{db}}} z_i^b = n^b
\label{eq:dc-blocker-count-cons}
\end{equation}
\noindent where
$n^b$ is the total number of blockers to be placed.

\vspace{0.1in}
\subsubsection{Cost of load-shedding} \label{subsubsec:bpc-cost-loadshed}
\indent

To restrict the total cost of load-shedding under a specified threshold, given that the system has loads with varying priorities:

\begin{equation}
\allowdisplaybreaks
\small
\smashoperator{\sum_{i \in \mathcal{D}}} \abs{p_i^d} \kappa_i^d z_i^d \le \overline{c}_d
\label{eq:load-shed-cons}
\end{equation}
\noindent where
$\overline{c}_d$ is the maximum allowable cost of load-shedding.

\vspace{0.1in}
\subsubsection{Fraction of load served} \label{subsubsec:bpc-fract-loadserved}
\indent

In the case, when the cost of blocker placement (Subsubsection~\ref{subsubsec:bpc-cost-blockers}) rather than load-shedding (Subsubsection~\ref{subsubsec:bpc-cost-loadshed}) is the quantity to be minimized, a supporting constraint may be the fraction of load served; this must be greater than a specified threshold:

\begin{equation}
\allowdisplaybreaks \footnotesize
\small
\smashoperator{\sum_{i \in \mathcal{D}}} \abs{p_i^d} z_i^d \le \rho_d \smashoperator{\sum_{i \in \mathcal{D}}} \abs{p_i^d}
\label{eq:fraction-load-served-cons}
\end{equation}
\noindent where
$\rho_d$ is the minimum fraction of load that must be served.

\section{Reference Implementation and Benchmarking}
\indent
\vspace{-0.1in}

The above introduced problem formulation was implemented in PowerModelsGMD.jl~v5.0, an open-source Julia tool within the InfrastructureModels software ecosystem \cite{multi-infrastructure}.
For the optimal placement of GIC dc-current blocking devices, the selected objective was to minimize the number of blockers placed -- therefore, minimizing the cost of blocker placement, as described in Subsubsection~\ref{subsubsec:bpc-cost-blockers} -- subject to Constraint~\eqref{eq:dc-blocker-const-cons}; additionally, the fraction of load served was used as supporting constraint -- load-shedding upper bounded was set to 15\% of total load, with all loads weighted equally, as described in Subsubsection~\ref{subsubsec:bpc-fract-loadserved} -- subject to Constraint~\eqref{eq:fraction-load-served-cons}.
This implementation is presented as a reference and is not intended for practical use on realistic network sizes; the intention is to quantify the computational difficulty of the blocker placement problem before heuristics are applied to scale to larger network sizes.

The implementation was benchmarked on a computer running Windows 10 with an Intel Xeon Silver 4112 CPU with 2.60 GHz and 64 GB RAM, using Juniper \cite{juniper} as a solver, and applied to three publicly available test cases of varying sizes (from 4 to 150 busses) \cite{pw_quick_start_guides, horton_test_2012, birchfield2016statistical}.
Results are summarized in Table~\ref{table:placement-benchmarks}: listing the optimal number of blockers placed (versus the total number of possible locations where blockers can be placed), the percentage of load met, the unit cost of the placed blocker(s), and the required computation time.

\begin{table}[!htbp]
\caption{Summary of GIC Blocker Placement Methods}
\begin{center}
\begin{tabular}{crrrrr}
\hline
Case & Busses & Blockers & Load Met & Cost & Runtime \\
& & & & &(CPU Seconds) \\ 
\Xhline{2\arrayrulewidth}
B4GIC & 4 & 1/2 & 100\% & 1.0 & 12.9 \\
EPRI21 & 20 & 1/8 & 81.8\% &  1.0
&  9.2\\
UIUC150 & 169 & 1/98 & 100\% & 1.0 & 1759 \\
\hline
\end{tabular}
\end{center}
\label{table:placement-benchmarks}
\vspace{-0.2in}
\end{table}

\section{Conclusion}
\indent
\vspace{-0.1in}

The threat of GMDs pose a risk to the BES, and addressing the challenges is difficult because of the lack of empirical data.
This means that simulation and optimization methods are important for analysis and impact mitigation.

This paper presented a complete base problem formulation for the optimal placement of transformer neutral (dc-current) blocking devices, an effective technology to reduce GICs without reducing the power-carrying capacity of the grid.
Although this formulation is not tractable -- as it is an NP-hard problem with mixed-integer nonlinear, nonconvex constraints -- and is formulated as an ac polar formulation -- other formulations, such as semidefinite or second-order cone programming \cite{coffrin_qc_2015}, may be used as well -- this is an important first step towards the development of a scalable, optimization-based method to guide the placement of blocking devices.

The outlined problem formulation was implemented in PowerModelsGMD.jl and benchmarked on three test cases of varying sizes, providing comparison for future work.
As the benchmarking experiments show, obtaining a globally optimum solution in reasonable time is increasingly difficult as the network size increases.
Relaxations and approximations -- for the GIC flow and $q_{loss}$ coupling constraints (including the absolute value, the $q_{loss}$ constraint, and the dc current nodal balance constraints) that all include bilinear terms -- need to be explored an applied to improve the computational performance and scale the base problem to realistic network sizes.
This is the subject of ongoing research work.


\bibliographystyle{unsrt}
\bibliography{references}

\begin{thebibliography}{10}

\bibitem{NA-resilience-2017}
{National Academies of Sciences, Engineering, and Medicine}.
\newblock {\em {Enhancing the Resilience of the Nation's Electricity System}}.
\newblock {The National Academies Press. Washington DC}, 2017.

\bibitem{NERC-HILF-2010}
{NERC} and {U.S. Department of Energy}.
\newblock {High-Impact, Low-Frequency Event Risk to the North American Bulk
  Power System}.
\newblock Technical report, {North American Electric Reliability Corporation},
  Jun. 2010.

\bibitem{mate21-pmsgmd-cascade}
A.~{Mate} et~al.
\newblock {Relaxation Based Modeling of GMD Induced Cascading Failures in
  PowerModelsGMD.jl}.
\newblock In {\em {Proceedings of the 2021 IEEE/PES 53rd North American Power
  Symposium}}, 2021.

\bibitem{mate21-pmsgmd}
A.~{Mate} et~al.
\newblock {Analyzing and Mitigating the Impacts of GMD and EMP Events on the
  Electrical Grid with PowerModelsGMD.jl}, Jan. 2021.
\newblock \url{https://arxiv.org/abs/2101.05042}.

\bibitem{juvekar2021gic}
G.~P.~{Juvekar} et~al.
\newblock {GIC-Inclusive State Estimator for Power System Awareness During
  Geomagnetic Disturbance Events}.
\newblock {\em {IEEE Transactions on Power Systems}}, 36(4):2966--2974, Jul.
  2021.

\bibitem{overbye2022towards}
T.~J.~{Overbye} et~al.
\newblock {Towards Developing Implementable High Altitude Electromagnetic Pulse
  E3 Mitigation Strategies for Large-Scale Electric Grids}.
\newblock In {\em {Proceedings of the 2022 IEEE Texas Power and Energy
  Conference}}, 2022.

\bibitem{kazerooni2017mitigation}
M.~{Kazerooni} et~al.
\newblock {Mitigation of Geomagnetically Induced Currents Using Corrective Line
  Switching}.
\newblock {\em {IEEE Transactions on Power Systems}}, 33(3):2563--2571, May
  2018.

\bibitem{lu2017optimal}
M.~{Lu} et~al.
\newblock {Optimal Transmission Line Switching Under Geomagnetic Disturbances}.
\newblock {\em {IEEE Transactions on Power Systems}}, 33(3):2539--2550, May
  2018.

\bibitem{ryu2022mitigating}
M.~{Ryu} et~al.
\newblock {Mitigating the Impacts of Uncertain Geomagnetic Disturbances on
  Electric Grids: A Distributionally Robust Optimization Approach}.
\newblock {\em {IEEE Transactions on Power Systems}}, 37(6):4258--4269, Nov.
  2022.

\bibitem{zhang2022machine}
R.~{Zhang} et~al.
\newblock {Machine Learning-Aided Enhancement of Power Grid Resilience to
  Electromagnetic Pulse Strikes}.
\newblock In {\em {Proceedings of the 2022 North American Power Symposium}},
  2022.

\bibitem{yang2019optimal}
P.-H.~{Yang} et~al.
\newblock {Optimal Placement of Grounding Small Resistance in Neutral Point for
  Restraining Voltage Fluctuation in Power Grid Caused by Geomagnetic Storm}.
\newblock {\em {IET Generation, Transmission \& Distribution}},
  13(8):1456--1465, Apr. 2019.

\bibitem{horton_magnetohydrodynamic_2017}
R.~{Horton}.
\newblock {Magnetohydrodynamic Electromagnetic Pulse Assessment of the
  Continental U.S. Electric Grid}.
\newblock Technical Report 3002009001, Electric Power Research Institute, 2017.

\bibitem{ning2019research}
X.~{Ning} et~al.
\newblock {Research on Optimal Placement for GIC Mitigation with Blocking
  Device}.
\newblock In {\em {IOP Conference Series: Materials Science and Engineering}},
  2019.

\bibitem{liang2019optimal}
Y.~{Liang} et~al.
\newblock {Optimal Blocking Device Placement for Geomagnetic Disturbance
  Mitigation}.
\newblock {\em {IEEE Transactions on Power Delivery}}, 34(6):2219--2231, Dec.
  2019.

\bibitem{zhu2014blocking}
H.~{Zhu} et~al.
\newblock {Blocking Device Placement for Mitigating the Effects of
  Geomagnetically Induced Currents}.
\newblock {\em IEEE Transactions on Power Systems}, 30(4):2081--2089, Jul.
  2014.

\bibitem{zhu2016blocking}
H.~{Zhu} et~al.
\newblock {Blocking Device Placement for Mitigating the Effects of
  Geomagnetically Induced Currents}.
\newblock In {\em {Proceedings of the 2016 IEEE Power and Energy Society
  General Meeting}}, 2016.

\bibitem{shengquan2022optimal}
W.~{Shengquan} et~al.
\newblock {Optimal Blocking Devices Placement for Geomagnetic Disturbance
  Mitigation Based on Sensitivity of Induced Geoelectric Fields}.
\newblock {\em {IEEE Access}}, 10:132814--132821, Nov. 2022.

\bibitem{lamichhane2022optimal}
A.~{Lamichhane}.
\newblock {\em {Optimal Mitigation of Geomagnetically Induced Current Effects
  in Power Systems Considering Transformer Thermal Limits}}.
\newblock {Masters Thesis}, York University, Dec. 2022.

\bibitem{etemadi2014optimal}
A.~H.~{Etemadi} et~al.
\newblock {Optimal Placement of GIC Blocking Devices for Geomagnetic
  Disturbance Mitigation}.
\newblock {\em {IEEE Transactions on Power Systems}}, 29(6):2753--2762, Mar.
  2014.

\bibitem{liang2015optimal}
Y.~{Liang} et~al.
\newblock {Optimal Blocker Placement for Mitigating the Effects of Geomagnetic
  Induced Currents Using Branch and Cut Algorithm}.
\newblock In {\em {Proceedings of the 2015 North American Power Symposium}},
  2015.

\bibitem{rezaei2018optimal}
A.~{Rezaei-Zare} et~al.
\newblock {Optimal Placement of GIC Blocking Devices Considering Equipment
  Thermal Limits and Power System Operation Constraints}.
\newblock {\em {IEEE Transactions on Power Delivery}}, 33(1):200--208, Feb.
  2018.

\bibitem{pirjola2000-gic}
R.~Pirjola.
\newblock {Geomagnetically Induced Currents During Magnetic Storms}.
\newblock {\em IEEE Plasma Sci.}, 28(6):1867--1873, Dec. 2000.

\bibitem{shetye2015-gmd}
K.~S.~{Shetye} et~al.
\newblock {Modeling and Analysis of GMD Effects on Power Systems: An Overview
  of the Impact on Large-Scale Power Systems}.
\newblock {\em IEEE Electrific. Mag.}, 3(4):13--21, Dec. 2015.

\bibitem{boteler2017modeling}
D.~H.~{Boteler} et~al.
\newblock {Modeling Geomagnetically Induced Currents}.
\newblock {\em {Space Weather}}, 15(1):258--276, Dec. 2016.

\bibitem{barnes2019resilient}
A.~{Barnes} et~al.
\newblock {Resilient Design of Large-Scale Distribution Feeders with Networked
  Microgrids}.
\newblock {\em Electric Power Systems Research}, 171(0):150--157, Jun. 2019.

\bibitem{overbye2012integration}
T.~J.~{Overbye} et~al.
\newblock {Integration of Geomagnetic Disturbance Modeling into the Power
  Flow}.
\newblock In {\em {Proceedings of the 2012 North American Power Symposium}},
  2012.

\bibitem{multi-infrastructure}
R.~{Bent} et~al.
\newblock {InfrastructureModels: Composable Multi-Infrastructure Optimization
  in Julia}.
\newblock 2023.
\newblock \textit{Under Review}.

\bibitem{juniper}
O.~{Kröger} et~al.
\newblock {Juniper: An Open-Source Nonlinear Branch-and-Bound Solver in Julia}.
\newblock In {\em {Proceedings of the Integration of Constraint Programming,
  Artificial Intelligence, and Operations Research}}, pages 377--386, 2018.

\bibitem{pw_quick_start_guides}
{PowerWorld Corporation}.
\newblock {Quick-Start Guides}.
\newblock \url{https://www.powerworld.com/training/quick-start-guides}, May
  2023.

\bibitem{horton_test_2012}
R.~{Horton} et~al.
\newblock {A Test Case for the Calculation of Geomagnetically Induced
  Currents}.
\newblock {\em {IEEE Transactions on Power Delivery}}, 27(4):2368--2373, Oct.
  2012.

\bibitem{birchfield2016statistical}
A.~{Birchfield} et~al.
\newblock {Statistical Considerations in the Creation of Realistic Synthetic
  Power Grids for Geomagnetic Disturbance Studies}.
\newblock {\em IEEE Transactions on Power Systems}, 32(2):1502--1510, Mar.
  2017.

\bibitem{coffrin_qc_2015}
C.~{Coffrin} et~al.
\newblock {The QC Relaxation: Theoretical and Computational Results on Optimal
  Power Flow}, Jul. 2015.
\newblock \url{http://arxiv.org/abs/1502.07847}.

\end{thebibliography}

\end{document}